\documentclass[nofootinbib,twocolumn,amsfonts,prd,aps]{revtex4}
\usepackage{graphicx}
\usepackage{subfig}
\usepackage{amsmath,amssymb,amsfonts,amsthm,mathrsfs}

\begin{document}

\title{Conformal loop quantum gravity coupled to the Standard Model}

\author{Miguel Campiglia$^{1}$, Rodolfo Gambini$^{1}$, 
Jorge Pullin$^{2}$}
\affiliation {
1. Instituto de F\'{\i}sica, Facultad de Ciencias, 
Igu\'a 4225, esq. Mataojo, 11400 Montevideo, Uruguay. \\
2. Department of Physics and Astronomy, Louisiana State University,
Baton Rouge, LA 70803-4001}

\begin{abstract}
  We argue that a conformally invariant extension of general
  relativity coupled to the Standard Model is the fundamental theory
  that needs to be quantized. We show that it can be treated by loop
  quantum gravity techniques. Through a gauge fixing and a modified
  Higgs mechanism particles acquire mass and one recovers general
  relativity coupled to the Standard Model. The theory
  suggests new views with respect to the definition of the Hamiltonian
  constraint in loop quantum gravity, the semi-classical limit and the
  issue of finite renormalization in quantum field theory in quantum
  space-time. It also gives hints about the elimination of ambiguities that
  arise in quantum field theory in quantum space-time in the
  calculation of back-reaction.

\end{abstract}

\maketitle

General relativity is not conformally invariant. This is due to the
presence of a dimensionful constant in the action, Newton's
constant. However, there exists a conformal extension of general
relativity that is locally conformally invariant and that is
equivalent to general relativity, as one can show by fixing a certain
family of gauges.  We will show that the theory can be coupled to
matter and defines a gravitational extension of the Standard Model
that is conformally invariant. Again one recovers the usual Standard
Model coupled to general relativity through a gauge fixing. It is
reasonable to think that a conformal extension of general relativity
is the fundamental theory that should be quantized. A strong
suggestion stems from the fact that it has been observed that the high
energy limit of all non-trivial renormalizable field 
theories is conformally invariant \cite{banks}.

But there are additional reasons to consider a conformal theory:
a) The quantization of gravity in terms of loops has encountered
important obstructions at the level of finding a quantization without
quantum anomalies. Particularly problematic is the implementation of
the Hamiltonian constraint that up to now has only allowed for an
ultralocal implementation \cite{marolf}. The main obstacle to a full
quantization is that the kinematics of loop quantum gravity is quite
limited at the time of implementing the constraints and making contact
with the usual non-diffeomorphism invariant semiclassical physical
picture of gravity. There have been some extensions proposed to deal
with this but none is still widely accepted (see for instance
\cite{dittrich,lewandowski,sahlmannthiemann,alok}).

b) Related to the last point, it has also been observed that the
existence of a Planck scale imposes restrictions at the process of
going to the continuum limit. The idea is that if one adds additional
points to the spin network in order to refine it, the continuum
approximation of volumes and areas does not improve and one just adds
volume to the space-time, as there is a minimum eigenvalue for
areas. Among other things, this makes it difficult to take the various
limits involved in the definition of the Hamiltonian constraint in a
non-trivial way. In a conformally invariant theory one does not have a
length scale and there is a chance to improve the situation concerning
these problems, as we shall see. 

c) Studies of quantum field theory in quantum space times defined by
loop quantum gravity have suggested that although the discreteness of
the quantum space-time makes the quantum field theory finite, a finite
renormalization is needed to remove dependence of the lower energy
physics with respect to the Planck scale degrees of freedom
\cite{nahuel,spherical}. Unfortunately, it appears that the amount of
renormalization depends on the state of the quantum space-time
background, in particular on the spacing of the vertices of its spin
networks. This is a potential problem, although we do not know whether
it will be present once the full interacting theory of gravitating
quantum fields will be fully quantized.

Conformal invariance can help with points a), b) and c): In a
conformal theory of gravity conformal spin networks can be defined
that can be indefinitely refined approximating the conformal geometry
with arbitrary precision. This suggests that conformal loop quantum
gravity will admit a formulation that may not present the problems of
the usual quantization. In particular one has excellent perspectives
that it will have a local Hamiltonian that is non-trivial as is
typical in discrete models.  And it opens
possibilities for the renormalization problems in quantum field theory
in quantum space-time since no counterterms are needed in the
renormalization that depend on spacing.

Let us outline the proposed theory.
It is well known that the Brans--Dicke theory with coupling
$\omega=-3/2$ is conformally invariant \cite{bd}. The action is given by,
\begin{equation}
  S = 3 \int d^4x \sqrt{-g}  \left[\frac{\phi^2 R}{6} + 
    g^{ab} \partial_a \phi \partial_b \phi\right]. \label{BDaction}
\end{equation}
In the gauge
$
\phi(x) =\kappa^{-1/2}, 
$
with $\kappa= 8 \pi G$ the theory is identical to Einstein's gravity. 
The combination
$
g^{(c)}_{ab}:= \phi^2 g_{ab}, 
$
is conformally invariant.  In terms of $g^{(c)}_{ab}$, the action (\ref{BDaction}) takes the manifestly conformally invariant form:
$
S= \frac{1}{2}\int d^4x \sqrt{-g^{(c)}} R^{(c)}. \label{CIaction}
$
It should be noted that the
conformal metric $g^{(c)}_{ab}$ has different dimensionality than the usual
metric. If one computes invariant intervals with the conformal metric
the result is dimensionless. 

For the loop quantization of this theory, we  will consider the Holst
version of the conformal action,
\begin{equation}
  S_H^{(c)} = \int \frac{1}{2} e^{(c)} e^{a(c)}_I e^{b(c)}_J \left(
\Omega_{ab}^{IJ(c)} +\frac{1}{\gamma} {}^*\Omega_{ab}^{IJ(c)}\right), \label{Holstaction}
\end{equation}
with $\gamma$ the Immirzi parameter and
${\Omega}_{ab}{}^{IJ (c)}$ the curvature of
the $SL(2,C)$ connection $\omega_a^{IJ (c)}$.
One can write this action in terms of the geometrical triad and connection by substituting 
\begin{eqnarray}
  e_a^{I(c)} &=& \phi e_a^I,\\
  e^{a(c)}_I &= & \frac{e^a_I}{\phi}, \\
  \omega^{IJ (c)}_a &= &  \omega^{IJ}_a -2 \phi^{-1} \partial_b \phi e^{b [I} e_a^{J]},
\end{eqnarray}
in (\ref{Holstaction}).
We now describe the relation between conformal and geometrical variables at the canonical level. 

First, we can express the theory  in Hamiltonian form in terms of (conformally invariant) triad and extrinsic curvature canonical pair:
\begin{equation}
\{K_a^{i (c)}(x), E^{b (c)}_j \}= \delta_a^b  \delta^i_j \delta^3(x,y),
\end{equation}
with a Hamiltonian, diffeomorphisms and Gauss constraints. Since the action is independent of the field $\phi$, its conjugated momentum vanishes. That is, we have the additional  canonical pair and constraint:
\begin{equation}
\{ \phi(x), \pi^{(c)}(y) \} = \delta^3(x,y),
\end{equation}
\begin{equation}
\pi^{(c)} \approx 0. \label{pic}
\end{equation}
The transformation to geometrical variables associated to the action (\ref{BDaction}) can be achieved through the canonical transformation:
\begin{eqnarray}
 E^{a}_i & = & \phi^{-2}E^{a (c)}_i, \\
 K^{i}_a & =& \kappa \, \phi^2 K^{i (c)}_a, \\
\pi & =&  \pi^{(c)}-2  \phi^{-1}  K^{i (c)}_a E^{a (c)}_i,
\end{eqnarray}
with $\phi$ unchanged. The new nonzero Poisson brackets
are 
\begin{equation} \{K_a^{i}(x), E^{b}_j \}= \kappa \, \delta_a^b
  \delta^i_j \delta^3(x,y), \quad \{ \phi(x), \pi(y) \}
  =\delta^3(x,y).  \end{equation} The constraint (\ref{pic}) becomes,
after a rescaling by $\phi$, the conformal
constraint: 
\begin{equation} \phi \, \pi^{(c)} = \frac{2}{\kappa} K_a^i E^a_i+ \pi
  \phi =: {\cal S}. \label{20}
\end{equation} 

Finally, in terms of Ashtekar
variables $A_a^i = \gamma K_a^i + \Gamma_a^i$, the remaining
Hamiltonian, diffeomorphisms and Gauss constraints can be written as:
\begin{eqnarray} 
H &=& \frac{\phi^2}{2} \epsilon_i^{lm} E^a_l
  E^b_m\left[F_{ab}^i-\left(\gamma^2 +(\kappa \phi^2)^{-2}
    \right)K_a^jK_b^k\epsilon^i_{jk}\right] \nonumber\\
&&- E^{ai} E^b_i \partial_a
  \phi \partial_b \phi + 2 E^{ai}E^b_i \left(\nabla_a \partial_b
    \phi\right)\phi,\\
{\cal C}_a
  &=&\frac{1}{\gamma} F_{ab}^iE^b_i+\pi \partial_a \phi,\\ 
  {\cal G}^i &=&\partial_a E^a_i +\epsilon_{ijk} A_a^j E^{ak}.\label{23}
\end{eqnarray}
It is clear that this theory is ready for a loop quantization.

We start by constructing the kinematical Hilbert space. As usual we
start by building variables that are gauge invariant 
given by parallel transports of the conformally invariant connection
$A_a^{i(c)}$, 
\begin{equation}
  U\left(A^{(c)},\eta\right)= P \exp \int A^{(c)}_a dy^a,
\end{equation}
with $\eta$ a path and $A^{(c)}_a$ the conformally invariant
connection
\begin{equation}
  A^{(c)}_a = \Gamma_a^{i(c)} + \gamma K_a^{i(c)},
\end{equation}
with $\Gamma_a^{i(c)}$ the conformal spin connection defined through
its usual relation with $E^{a(c)}_i$ (in particular, it has the same SU(2) transformation properties as the Ashtekar connection $A_a^i$).

Under gauge transformations the holonomies
transform as 
\begin{equation}
  U\left(A^{(c)},\eta\right)\to \Lambda\left(x_f^\eta\right)
U\left(A^{(c)},\eta\right) \Lambda^{-1}\left(x_i^\eta\right)
\end{equation}
with $x_f^\eta$ and $x_i^\eta$ the ending and starting points of the
path $\eta$ and $\Lambda$'s are finite gauge transformation
matrices. The conformal holonomy has vanishing Poisson bracket with
the conformal constraint. In terms of them one can define conformally
invariant spin networks in the usual way. The cylindrical conformal
functions are defined as,
\begin{equation}
  \psi_S\left(A^{(c)}\right) = \otimes_l
  R^{(j_l)}\left(U\left(A^{(c)},\eta_l\right)\right)
\otimes_n i_n,
\end{equation}
where $R^{(j_l)}$ is a representation of the group, in the case of
pure gravity
$SU(2)$, of dimension $2j+1$, $l$ is the label of the path, and $i_n$
are the intertwiners associated to the vertex $v_n$. These spin
networks are trivially annihilated by the quantum version of 
Gauss' law, and the conformal constraint (\ref{20}). Through
group averaging can be made diffeomorphism invariant. We will not
expand on that since it is the same construction as in the usual
theory. In particular the orthogonality properties of the conformally
invariant spin knots $\vert S_{(c)}\rangle$ are as in the usual case.
One can define conformal operators in the basis $\vert S^{(c)}\rangle$
that characterize properties of a conformal manifold, in particular
angular properties are invariant under conformal transformations. 

For many situations the conformally invariant spin nets can be defined
on a compact spatial manifold. For instance if one considers asymptotically
flat space-times infinity can be brought to a finite boundary via a
conformal transformation. The compact manifold is a Penrose diagram.

The geometric operators corresponding to the area of a surface 
and the volume of a region can be extended to conformal
operators substituting in the expressions that define the
geometric operators $g, E, A$ with their conformally invariant
counterparts $g^{(c)}, E^{(c)}, A^{(c)}$. The spectrum of the
conformally invariant areas and volumes are the same as the usual ones
with the exception that the factors involving the Planck area and
volume do not appear. These operators define the conformal geometry
associated to the spin net $\vert S^{(c)}\rangle$. 

Something different from the usual theory is that in the conformal
case one can indefinitely add vertices to the spin network to refine
it without any limitations in order to approximate a given conformal
geometry. In the non-conformal case if one adds vertices the volume of
the region increases and therefore the geometry changes. This limits
how well one can approximate a given geometry.

It is possible to couple massless matter to conformal gravity. In fact
one can couple the complete Standard Model through a mechanism in
which the Higgs boson acquires mass in the gauge fixed conformal
theory. When one gauge fixes, the Planck scale becomes determined and from
there one generates the Higgs mass and indirectly the masses of all
known elementary particles. We will see that the relation between
their masses and the Planck mass will be given by the dimensionless
constants of the theory. 

It will be possible to rewrite the total Lagrangian as follows,
\begin{equation}
  {\cal L}_T = {\cal L}_{\rm GR}\left(g^{(c)}\right) +{\cal
      L}_M\left(g^{(c)}, \frac{\Psi^M}{\phi^d}\right),
\end{equation}
where the $\Psi^M$ are the matter fields and $d$ is a suitable power
of $\phi$ that is introduced to ensure conformal invariance of the
matter fields, as we shall see. The equations of motion imply that the
stress tensor of the matter fields is traceless. With this one can
incorporate all the particles in the Standard Model, without mass. The
only thing that requires special discussion is the Higgs field, which
endows all other fields with mass via the Higgs mechanism.

Let us recall that in the Standard Model one considers a Higgs boson
in a given representation, we will take here the fundamental one where
$H^\alpha(x)$ is a doublet, with $\alpha=1,2$. Let us focus on the
portion of the Lagrangian that involves $H^\alpha(x)$,
\begin{eqnarray}
  {\cal L}_H &=&\!\!\!
\left[-\frac{1}{2} \left( \partial_a H^{\dagger\alpha}+ g A_a^i
      \tau^{i\alpha}_\beta H^{\dagger\beta}\right)\left(\partial^a H^\alpha +g
      A^{ak}\tau_{k\gamma}^\alpha
      H^\gamma\right)\right.\nonumber\\&&\left.
-V\left(H\right) 
\right]\sqrt{-g},\label{mexican}
\end{eqnarray}
with the potential given by,
\begin{equation}
  V\left(H\right)= \frac{\lambda}{4}\left(H^{\dagger\alpha}H^\alpha\right)^2-\mu^2
  H^{\dagger\alpha} H^\alpha+ {\rm const.}
\end{equation}
which corresponds to a term of mass $-\mu^2$ that gives rise to a
Mexican hat potential. In the above expression we assume that $A_a^i$ 
is a Yang--Mills connection associated with the weak interactions $\tau$
are the generators of $SU(2)$, $\lambda$ is a dimensionless coupling
constant. For simplicity, we are not including in (\ref{mexican}) 
the $U(1)$ boson field.

One can construct loop invariants that include matter
in terms of $\Psi^{(c)} =\Psi/\phi^{3/2}$ $H^{\alpha(c)} =
H^\alpha/\phi$. The fermions can be included at the ends of open
paths, the scalar bosons can be included anywhere. The spin networks
including matter will have valences in each path not only associated
with the representation of the $SU(2)$ associated with gravity, but
also with the representations of all the vector bosons \cite{rovelli}

To consider the loop framework it will be good to write the action in
terms of conformal variables. In particular, for the portion involving
the Higgs field, it reads,
\begin{eqnarray}
  S_{\rm Higgs} &=& \int d^4x \sqrt{-g_{(c)}}\left[ -g^{ab(c)} D_a H^{(c)\dagger} D_b
  H^{(c)}  \right.\nonumber\\
&&-\frac{\lambda}{4}
  \left( H^{(c)\dagger} H^{(c)} -\alpha^2\right)^2 +\frac{\lambda'}{4}
  \nonumber\\
&&\left.+{\cal L}_{SM}\left(g^{(c)},\Psi^{(c)}, A_a\right)\right]. 
\end{eqnarray}
For brevity we do not list the Standard Model terms explicitly, but
one can easily show that the standard action for the Dirac fields in
curved space time  may be written without any
change of form in terms of the conformal invariant combinations as in (21).

Since the above action is written entirely in terms of variables that
are conformally invariant, the term involving one sixth the Higgs
field squared times the scalar curvature is not needed to
enforce conformal invariance \cite{bars} If added it
would would lead to a slightly different particle physics only in
situations where gravity is important.  When curvature is negligible
with the gauge fixing $\phi=\kappa^{-1/2}$ it coincides with the usual
Standard Model.

If one considers a gauge fixing $\phi(x)=\phi_0={\rm constant}$ one
can write the dimensionful parameters in terms of $\phi_0$, 
\begin{eqnarray}
G  &=& \frac{1}{8\pi \phi_0^2},\\
\frac{\Lambda}{16\pi G}&=& \frac{\lambda' \phi_0^4}{4},
\end{eqnarray}
with $\Lambda$ the cosmological constant. This is reminiscent of how
Newton's constant emerges in string theory \cite{zwiebach}. 
If one now carries out the
Higgs mechanism for the resulting theory, one gets for the mass and
the expectation value of the Higgs,
\begin{eqnarray}
\langle H^{\dagger}H\rangle &=& \alpha^2 \phi_0^2,\\
  m_{\rm Higgs}^2 &=& \lambda \alpha^2 \phi_0^2.
\end{eqnarray}
A similar construction can be done
for all the other massive particles in the Standard Model. Notice that
all the masses and expectation values are determined in terms of the
Planck scale and dimensionless parameters, no matter what choice of
gauge fixing, implying that the physics is gauge invariant.

Physics beyond the Standard Model, like neutrino oscillations,
could also be incorporated in the conformal context through mechanisms
like the one in \cite{nicolai}.

Given that the gauge fixed theory is the usual one, it could be asked
what has been gained by introducing the conformal theory.  First of
all, the quantization of a gauge fixed theory is usually inequivalent
to the Dirac quantization of the full theory. However, the most
interesting answer is that in a quantum theory the constants are
expected to be running coupling constants. If one chooses the Planck
scale $G$ fixed, the relations of the masses of particles in terms of
the Planck mass could therefore change. In the low energy world we
live in the particle masses have a fixed relation to the Planck
mass. This would change at high energies where quantum gravitational
effects are important. At that level the best way to depict things is
conformally invariant and therefore this suggests that the conformal
theory is the correct theory to quantize. An alternative proposal is
the one by 't Hooft \cite{hooft}. He noted that in conformally
invariant theories there is a contribution from dimensional
regularization that depends on the dilaton field, which suggests that
to keep conformal invariance at a quantum level the dimensionless
constants are determined by requiring that the beta function of the
renormalization group vanishes. In either case the theory makes
non-trivial predictions.

Another advantage of the conformal theory arises when one considers
the quantization of a field theory living in a quantum space-time as
for instance in \cite{hawking} in which the background is provided by
a loop quantization of the space-time. The divergent
terms of the stress energy tensor depend on the quantum state that
represents the background space-time \cite{nahuel,spherical}. That is
clearly not acceptable since to have low energy physics be independent
of such state one would have to include counterterms that are
state-dependent. This is not possible in the theory with fixed $G$. It
is possible in the conformal theory. In that theory one can absorb the
terms that would give divergences in the cosmological constant and the
curvature in the action in the continuum theory simply by changing the
gauge. This also suggests a solution to the problem noted by Wald
\cite{wald} that an ambiguity appears in the renormalization for the
terms quadratic in the curvature. They depend on constants not
determined by the theory and therefore there is ambiguity in computing
back reaction. In the conformal theory the terms quadratic in the
curvature cannot get corrections since they are independent of
Newton's constant. Thus, if one assumes that these terms vanish in the
original action, in a finite theory like the one stemming from a loop
quantization, they would emerge from the terms that diverge
logarithmically in the continuum. These would just provide very small
corrections to the action that can be viewed as originating in quantum
gravity effects. Since these terms are not reabsorbed, the conformal (trace)
anomaly of quantum field theory in curved space time would not seem to
appear.

As we have introduced an extra degree of freedom and an extra
constraint, one may think that one has introduced a spurious
invariance that does not add anything to the usual description
\cite{jackiw}.  However, at the quantum level, \cite{hooft,banks} as
we have already noticed, the conformal treatment could have deep
consequences in the analysis of quantum anomalies even ignoring its
implications on the loop quantization of gravity The introduction of
the dilaton also opens the possibility of having matter fields only
coupled to gravity in a conformal invariant way. For instance one
could consider spinors coupled to the dilaton via Yukawa couplings,
which would allow to have purely gravitating fermions of arbitrary
mass. These could be dark matter candidates.

We have presented a conformally invariant theory of gravity coupled to
the Standard Model that is amenable to the quantization techniques of
loop quantum gravity.  The values of the dimensionful parameters are
all determined in terms of the Planck scale and dimensionless
parameters.  Kinematically its Hilbert space is given by conformal
spin networks with edges labeled with the representations of all the
gauge groups involved in gravity and the Standard Model. Dynamically
one has the expectation that there could be improvements in the
treatment of the Hamiltonian constraint. Because the theory admits
infinite refinements of the spin networks this opens the possibility
of finding a Hamiltonian constraint whose action would not be
ultralocal.

This work was supported in part by Grant
No. NSF-PHY-1305000, NSF-PHY-1603630, ANII
FCE-1-2014-1-103974, funds of the Hearne Institute for Theoretical
Physics, CCT-LSU, FQXi, and Pedeciba.

\end{document}